\documentclass[aps,floatfix,twocolumn]{revtex4}
\usepackage{graphics,epsfig}
\usepackage{graphicx}

\begin{document}

\title{The flatness problem and $\Lambda$}
\author{Kayll Lake \cite{email}}
\affiliation{Department of Physics, Queen's University, Kingston,
Ontario, Canada, K7L 3N6 }
\date{\today}
\begin{abstract}
By way of a complete integration of the Friedmann equations, in
terms of observables, it is shown that for the cosmological
constant  $\Lambda > 0$ there exist non-flat FLRW models for which
the total density parameter $\Omega$ remains $\sim 1$ throughout
the entire history of the universe. Further, it is shown that in a
precise quantitative sense these models are not finely tuned. When
observations are brought to bear on the theory, and in particular
the WMAP observations, they confirm that we live in just such a
universe. The conclusion holds when the classical notion of
$\Lambda$ is extended to dark energy.
\end{abstract}
\maketitle

\begin{figure}[ht]
\epsfig{file=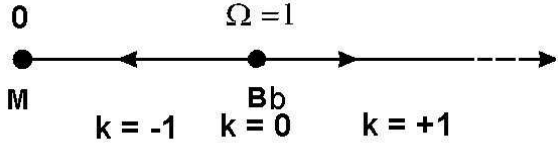,height=1in,width=3in,angle=0}
\caption{\label{state1}The state space for a single fluid FLRW
model with $p=(\gamma-1)\rho$ and $\gamma > 2/3$. \textbf{M} is
the Milne universe (Minkowski space). \textbf{Bb} is the spatially
flat FLRW model and the Big Bang. Turning points ($H=0$) are sent
to infinity by the definition (\ref{omega1}). The diagram reflects
the evolution for expanding universes.}
\end{figure}

The flatness problem is often considered to be the most impressive
issue in standard cosmology that is addressed by the inflation
paradigm \cite{guth}. Let us start by summarizing the flatness
problem for zero cosmological constant. With $\Lambda=0$ the
density parameter of a FLRW model is given by \cite{notation}
\begin{equation}
\Omega=\frac{8 \pi G \rho}{3 H^2 c^2} \label{omega1}
\end{equation}
and for a single fluid the state space is summarized for standard
models in FIG.~\ref{state1} \cite{ellis1}.

The essential point is that
except for the spatially flat case
\begin{equation}
\Omega=\Omega(t). \label{omega2}
\end{equation}
Observations show that
\begin{equation}
\Omega_{0} \sim 1. \label{omega3}
\end{equation}
The flatness problem involves the explanation of (\ref{omega3})
given (\ref{omega2}). The problem can be viewed in two ways.
First, one can take the view that there is a tuning problem in the
sense that at early times $\Omega$ must be finely tuned to $1$
\cite{peebles}. However, this argument is not entirely convincing
since \textit{all} standard models necessarily start with $\Omega$
exactly $1$. More convincing is the view that except for the
spatially flat case the probability that $\Omega \sim 1$ is
strongly dependent on the time of observation \cite{ellis2} and so
there is an epoch problem: why should (\ref{omega3}) hold?

\bigskip

In this letter we point out that if $\Lambda > 0$ then there exist
standard models for which $\Omega \sim 1$ throughout their entire
evolution even though they are not spatially flat. Moreover, in a
precise quantitative sense, we show that these models are
\textit{not} finely tuned. When current observations are brought
to bear on the theory, they confirm that we live in just such a
universe \cite{comp}.

\bigskip

\begin{figure}[ht]
\epsfig{file=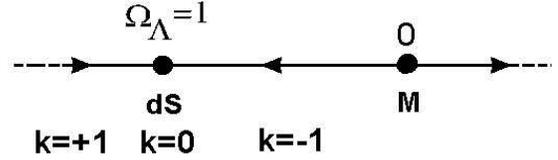,height=1 in,width=3in,angle=0}
\caption{\label{state2}The state space for $\Omega_{M}=0$ (here
rotated left $\pi/2$ to save space). Again \textbf{M} is the Milne
universe (Minkowski space \cite{specialrw}). \textbf{dS} is the
spatially flat representation of de Sitter space and acts as an
attractor for the two other representations of de Sitter space.}
\end{figure}

To include $\Lambda$ with dust define, in the usual way,
\begin{equation}
\Omega_{\Lambda} \equiv \frac{c^2 \Lambda}{3 H^2},\;\;\;\Omega_{k}
\equiv -\frac{c^2 k}{H^2 R^2}, \label{omegas}
\end{equation}
so that the Friedmann equations reduce to
\begin{equation}
\Omega_{M}+\Omega_{\Lambda}+\Omega_{k}=1 \label{omegatotal}
\end{equation}
where we have written $\Omega=\Omega_{M}$ for convenience. To
complement FIG.~\ref{state1}, which is the state space for
$\Omega_{\Lambda}=0$, the state space for $\Omega_{M}=0$ is shown
in FIG.~\ref{state2}.

Evolution in the full $\Omega_{\Lambda} - \Omega_{M}$ plane was
first considered by Stabell and Refsdal \cite{stabell} for the
case of dust ($\gamma=1$). Writing $Z=1+z$ ($1+z$ the standard
redshift), they populated the phase plane from relations
equivalent to
\begin{equation}
\Omega_{M}=\frac{\Omega_{M_{o}}Z^3}{\Omega_{M_{o}}Z^3+(1-\Omega_{M_{o}}-\Omega_{\Lambda_{o}})Z^2+\Omega_{\Lambda_{o}}},
\label{stabell1}
\end{equation}
which is a restatement of the conservation law for dust (the
constancy of $\Omega_{M}H^2R^3$), and
\begin{equation}
\Omega_{\Lambda}=\frac{\Omega_{\Lambda_{o}}}{\Omega_{M_{o}}Z^3+(1-\Omega_{M_{o}}-\Omega_{\Lambda_{o}})Z^2+\Omega_{\Lambda_{o}}},
\label{stabell2}
\end{equation}
which is a restatement of the constancy of $\Lambda$. They
distinguished trajectories via (half) the associated absolute
horizon, $\int_{0}^{\infty}\frac{cdt}{R}$. The same technique has
been applied more recently and in more general situations
\cite{phase}.

\bigskip

Equivalently, from a dynamical systems point of view
\cite{dynamics}, we can, again using dust as an example, consider
the system of differential equations
\begin{equation}
\Omega_{\Lambda}^{'}=(\Omega_{M}-2\Omega_{\Lambda}+2)\Omega_{\Lambda},
\label{system1}
\end{equation}
and
\begin{equation}
\Omega_{M}^{'}=(\Omega_{M}-2\Omega_{\Lambda}-1)\Omega_{M},
\label{system2}
\end{equation}
where $^{'} \equiv d/d\eta$ and $\eta \equiv ln(1/Z)$. We now
recognize the critical points in FIG.~\ref{state1} and
FIG.~\ref{state2}: \textbf{Bb} is a repulsor, \textbf{dS} an
attractor and \textbf{M} a saddle point for the full
$\Omega_{\Lambda}-\Omega_{M}$ plane.

\bigskip

\begin{figure}[ht]
\epsfig{file=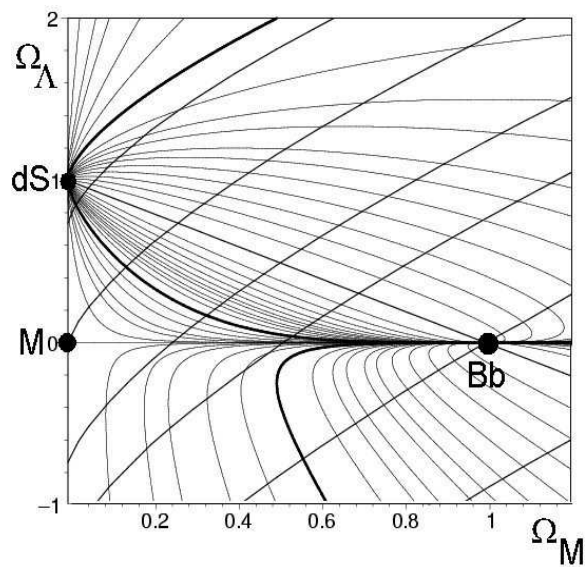,height=3in,width=3in,angle=0}
\caption{\label{alphalocus}The loci $\alpha = constant$ and
$Ht=constant$ \cite{age} in the $\Omega_{\Lambda} - \Omega_{M}$
plane. The loci shown are (bottom up) $Ht =
0.6,\;2/3,\;3/4,\;0.826,\;1,\;1.5,\;\infty\;\; (\equiv \alpha=1)$
(thick) and $| \alpha |
=0.01,\;0.05,\;0.1,\;1/4,\;1/2,\;1\;$(thick),
$2,\;3,\;4,\;6,\;8,\;12,\;20,\;40,\;100,\;500$. Note that for
$\Lambda > 0$ the delimiter $\Omega_{\Lambda}=1-\Omega_{M}$
($k=0$) is approached on either side ($k= \pm 1$) as $\alpha
\rightarrow \infty$. }
\end{figure}

The approach used here is somewhat different. Although what now
follows can be generalized \cite{further}, we continue with dust
as it presents an uncluttered relevant example. We observe a
constant of the motion ($\alpha$) for the system
(\ref{system1})-(\ref{system2}). The constant is given by
\cite{form}
\begin{equation}
(\frac{\Omega_{M}}{2 \Omega_{\Lambda}})^2 \pm \alpha
(\frac{\Omega_{k}}{3 \Omega_{\Lambda}})^3 =0 \label{lambdaelocus}
\end{equation}
for $k = \pm 1$ respectively. Since
$\Omega_{\Lambda}=1-\Omega_{M}$ for $k=0$, the system can be
considered solved and the Friedmann equations, in terms of
observables, in effect integrated. A phase portrait with
trajectories distinguished by $\alpha$ is shown in
FIG.~\ref{alphalocus}.

\bigskip

The physical meaning of $\alpha$ is not hard to find. In terms of
the bare Friedmann equation
\begin{equation}
\dot{R}^2+k=\frac{C}{R}+\frac{\Lambda R^2}{3}, \label{friedmann1}
\end{equation}
with ($k,\Lambda$) considered given, each value of the constant
$C>0$ determines a unique expanding universe ($k \neq 0$). For
each $C$, if $k=1$, there is a special value of $\Lambda$ that
gives a static solution (the Einstein static universe or
asymptotic thereto),
\begin{equation}
\Lambda_{E}=\frac{4}{9C^2}. \label{einstein}
\end{equation}
In terms of $\Lambda_{E}$ we have
\begin{equation}
\alpha=\frac{\Lambda}{\Lambda_{E}} \label{alpha}
\end{equation}
and so $\alpha$ is a measure of $\Lambda$ relative to the
``Einstein" value \cite{static}. The state spaces shown in
FIG.~\ref{state1} and FIG.~\ref{state2} correspond to $\alpha=0$:
$\Lambda=0$ in the first case and $C=0$ in the second.

\bigskip

As FIG.~\ref{alphalocus} makes clear, if $\Lambda>0$ and $k \neq
0$ then
\begin{equation}
\Omega \equiv \Omega_{M}+\Omega_{\Lambda} \sim 1 \label{flat}
\end{equation}
throughout the entire evolution of the associated universe if $
\alpha \in (a,\infty) $ where $a$ is a matter of choice but of the
order $> \sim 500$. In any event, it is clear that $\alpha$ need
not be finely tuned to produce (\ref{flat}) \cite{zero}.

\bigskip

\begin{figure}[ht]
\epsfig{file=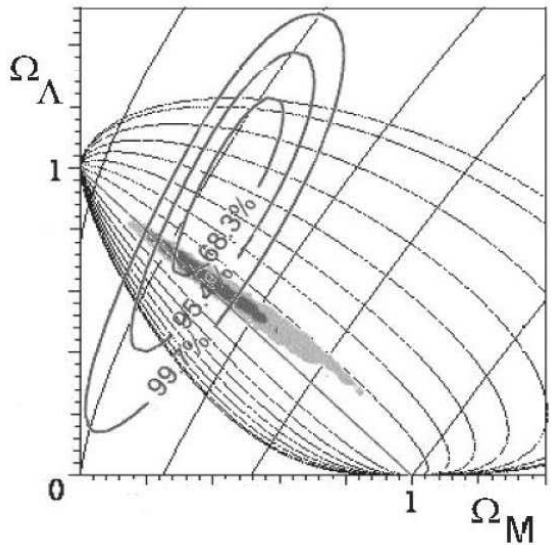,height=3in,width=3in,angle=0}
\caption{\label{alphadata} \textit{WMAP}  \cite{wmap}
 and \textit{HST} results \cite{Riess} superimposed on part of the
 phase portrait of FIG.~\ref{alphalocus}. In the
former case the \textit{``WMAP only"} results are used and in the
latter case earlier confidence levels have been removed for
clarity. The values of $H_{0}t_{0}$ shown are (top down) 1.5, 1,
0.826, 3/4 and 2/3. Above the delimiter
$\Omega_{\Lambda_{0}}=1-\Omega_{M_{0}}$, $k=1$ and the values of
$\alpha$ shown are (bottom up) 500, 100, 40, 20, 12, 8, 6 and 5.4.
Below the delimiter $k=-1$ and the values of $\alpha$ shown are
(top down) 500, 100, 40, 20, 12, 8, 6 and 5.4.
 }
\end{figure}

Fortunate we are to live in an era of unprecedented advances in
observational cosmology. We now bring some of theses observations
to bear on the foregoing discussion. The Wilkinson Microwave
Anisotropy Probe (\textit{WMAP}) has enabled accurate testing of
cosmological models based on anisotropies of the background
radiation \cite{wmap}. Independently, the recent Hubble Space
Telescope (\textit{HST}) type Ia supernova observations
\cite{Riess} not only confirm earlier reports that we live in an
accelerating universe \cite{Riess98} \cite{Perlmutter}, but also
explicitly sample the transition from deceleration to
acceleration. A comparison of these results in a partial phase
plane is shown in FIG.~\ref{alphadata} \cite{current}. Clearly the
(\textit{WMAP}) results, and to a somewhat lesser extent the
(\textit{HST}) results \cite{furtherhst}, support the view that we
live in ``large" $\alpha$ universe for which (\ref{flat}) has held
throughout its entire evolution.

\bigskip

We have argued that there are an infinity of standard non-flat
FLRW dust models for which $\Omega \sim 1$ throughout their entire
evolution as long as $\Lambda > 0$. Further, we have shown that
the WMAP observations confirm that we live in just such a
universe. The idea can be generalized in a straightforward way to
more sophisticated multi-component models wherein  $\Omega \sim 1$
($\Omega=\Sigma_{i} \Omega_{i}$ for all species $i$ but not
$\Omega_{k}$) from \textbf{Bb} to \textbf{dS} within a phase
portrait hyper-tube about $k=0$ (see \cite{further} for a
two-component model). Note, however, that the dust model examined
here is an excellent approximation now and the model relevant to a
comparison with current observations. With $\Omega \sim 1$ there
is no tuning or epoch problem and so no flatness problem in the
traditional sense. However, our presence in this hyper-tube is
presumably favored and an explanation of this probability about
the $k=0$ trajectory in a sense presents a refinement of the
classical flatness problem.

\bigskip

Since the analysis here has been based on the classical notion of
$\Lambda$, it is appropriate to conclude with a query as to
whether or not this analysis is stable under perturbations in the
definition of $\Lambda$ itself (that is ``dark energy" as opposed
to the classical notion of $\Lambda$). Since $\Lambda$ can be
introduced by way of a component $p_{w}=w \rho_{w}$ where $w
\equiv -1$, consider $w$ unspecified (but $<-1/3$). It can then be
shown that the analysis given here generalizes in a
straightforward way \cite{wcalc} and that for perturbations about
$w=-1$ the phase portrait given in FIG.~\ref{alphalocus} is
stable, but the state space for $\Omega_{M}=0$ shown in
FIG.~\ref{state2} is not \cite{state2pert}. This latter point in
no way alters the fact that there remain an infinity of non-flat
FLRW models for which $\Omega \sim 1$ throughout their entire
evolution as long as $W (\equiv 8 \pi \rho_{w}R^{3(1+w)})> 0$.

\bigskip

\begin{acknowledgments}
It is a pleasure to thank Phillip Helbig, James Overduin, Sjur
Refsdal and Rolf Stabell for their comments. This work was
supported by a grant from the Natural Sciences and Engineering
Research Council of Canada.
\end{acknowledgments}


\begin{thebibliography}{}
\bibitem{guth}See, for example, A. H. Guth, Physics Reports
\textbf{333-334}, 555 (2000). Note that the argument given there
is claimed to hold for $\Lambda \neq 0$, but as the argument given
here shows, it does not.
\bibitem{specialrw}There are precisely six empty Robertson-Walker
metrics: de Sitter space, which can be given in three forms ($k =
\pm1,0$); Minkowski space, which can be given in two forms
($k=-1,0$); and anti de Sitter space, which can be given only in
the form $k=-1$. All but the flat form of Minkowski space are
shown in FIG.~\ref{state2}, the flat form being sent to infinity
via the definition (\ref{omegas}).
\bibitem{static}The constant $\alpha$ has this physical
interpretation only for $k=+1$ and $\Lambda > 0$ but we make use
of the same constant of motion for $k=-1$ and, in
FIG.~\ref{alphalocus}, also for $\Lambda <0$.
\bibitem{current} The observations of course involve the current
values ($\Omega_{\Lambda_{o}},\Omega_{M_{o}}$). Our position in
the diagram corresponds to the intersection of the ``correct"
value of $H_{0}t_{0}$ with the correct trajectory determined
uniquely by $\alpha$ modulo $k$.
\bibitem{wcalc}Introducing the (non-interacting) species $p_{w}=w
\rho_{w}$ ($w<-1/3$) with the dust and with no concern here for
the classical energy conditions, we have the constant $W \equiv 8
\pi \rho_{w}R^{3(1+w)}$ and so define $\Omega_{W} \equiv
c^2W/3H^2R^{3(1+w)}$ to replace $\Omega_{\Lambda}$. It follows
that equations (\ref{stabell1}), (\ref{stabell2}), (\ref{system1})
and (\ref{system2}) are replaced by
\begin{eqnarray}
\Omega_{M}=\frac{\Omega_{M_{o}}Z^3}{\Omega_{M_{o}}Z^3+(1-\Omega_{M_{o}}-\Omega_{W_{o}})Z^2+\Omega_{W_{o}}Z^{3(1+w)}},\nonumber
\end{eqnarray}
\begin{eqnarray}
\Omega_{W}=\frac{\Omega_{W_{o}}Z^{3(1+w)}}{\Omega_{M_{o}}Z^3+(1-\Omega_{M_{o}}-\Omega_{W_{o}})Z^2+\Omega_{W_{o}}Z^{3(1+w)}},\nonumber
\end{eqnarray}
\begin{eqnarray}
\Omega_{W}^{'}=(\Omega_{M}+(1+3w)\Omega_{W}-(1+3w))\Omega_{W},\nonumber
\end{eqnarray}
and
\begin{eqnarray}
\Omega_{M}^{'}=(\Omega_{M}+(1+3w)\Omega_{W}-1)\Omega_{M}\nonumber
\end{eqnarray}
respectively. More importantly, (\ref{lambdaelocus}) generalizes
to
\begin{eqnarray}
3(\frac{\Omega_{M}}{\Omega_{W}})^{-1-3w} +
\frac{\alpha}{w}(\frac{3 w}{1+3 w})^{1+3 w} (\frac{\Omega_{W}}{\mp
\Omega_{k}})^{3w} =0\nonumber
\end{eqnarray}
for $k=\pm 1$ respectively. Now $\alpha=W/W_{E}$ where
\begin{eqnarray}
W_{E} = -\frac{1}{w}(\frac{3 w C}{1+3w})^{1+3w}.\nonumber
\end{eqnarray}
The stability of the phase portrait under perturbations on $w$ now
follows directly.
\end{thebibliography}
\end{document}